\documentclass[aps,amsfonts,prl,twocolumn,showpacs]{revtex4-1}

\usepackage{pdfpages} 
\makeatletter
\AtBeginDocument{\let\LS@rot\@undefined}
\makeatother

\usepackage{subfigure}
\usepackage{textcomp}
\usepackage{graphicx}
\usepackage{bm}
\usepackage{amsmath,amssymb,amsthm}
\usepackage{amsfonts}
\usepackage{dsfont}
\usepackage[colorlinks=true,linkcolor=blue,urlcolor=blue,citecolor=blue]{hyperref}
\usepackage{multirow}
\usepackage{url}

\def\parfig#1#2{
\parbox{#1\textwidth}
{\includegraphics[width=#1\textwidth]{#2}}
}

\begin{document}
\title{Tunable superdiffusion in integrable spin chains using correlated initial states}

\author{Hansveer Singh$^{1}$, Michael H. Kolodrubetz$^{3}$, Sarang Gopalakrishnan$^{2}$ and Romain Vasseur$^{1}$}
\affiliation{$^1$Department of Physics, University of Massachusetts, Amherst, Massachusetts 01003, USA \\
$^2$Department of Electrical and Computer Engineering, Princeton University, Princeton NJ 08544, USA\\
$^3$Department of Physics, The University of Texas at Dallas, Richardson, Texas 75080, USA}

\begin{abstract}
Although integrable spin chains only host ballistically propagating particles they can still feature diffusive spin transport. This diffusive spin transport originates from quasiparticle charge fluctuations inherited from the initial state's magnetization Gaussian fluctuations. We show that ensembles of initial states with quasi-long range correlations lead to superdiffusive spin transport with a tunable dynamical exponent. We substantiate our prediction with numerical simulations and explain how deviations arise from finite time and finite size effects.
\end{abstract}

\maketitle
\emph{Introduction}---In lattice quantum systems, typical initial states far from equilibrium relax according to diffusive hydrodynamics. After a local equilibration step, one expects that the only data from the initial state that survive in local observables are those which determine conjugate thermodynamic quantities, e.g. local temperature, chemical potential. The remaining evolution from local to global equilibrium is governed by the classical theory of hydrodynamics, which generically predicts diffusive transport for lattice systems.  However this intuition breaks down in kinetically constrained dynamics~\cite{GARRAHAN2018130,Fredrickson:1984tl,https://doi.org/10.48550/arxiv.1009.6113,Ritort_2003,Everest:2016wv, Iaconis:2019vd,Feldmeier:2020tq,Gromov:2020vu,Morningstar:2020tw, PhysRevX.10.011042,sala2020ergodicity,khemani2020localization,Ljubotina:2023aa} which can exhibit both sub- and superdiffusion, or avoid thermalization altogether~\cite{Moudgalya_2022}, and in integrable systems where initial state fluctuations in densities of infinitely-many conserved quantities play a prominent role in governing hydrodynamics.\par

	In integrable systems one has an extensive number of extensive conserved quantities or, equivalently, stable ballistically propagating quasiparticles at finite energy density~\cite{PhysRevX.6.041065,PhysRevLett.117.207201,PhysRevLett.119.220604,Bastianello_2022}. Although such systems are in principle fine tuned, they are of considerable interest since current experiments can engineer systems that are approximately integrable~\cite{Kinoshita_2006,Scheie_2021,Zundel:2019aa,Hild:2014aa,doi:10.1126/science.abk2397,doi:10.1126/science.1224953,Jepsen_2020,Tang:2018aa,Erne_2018}. Naively, stable ballistically-propagating quasiparticles should always lead to ballistic transport.
However this does not always hold, as highlighted by the $|\Delta| > 1$ regime of the spin-1/2 XXZ chain where one finds diffusive spin transport~\cite{Damle:1998aa,PhysRevLett.78.943,PhysRevLett.121.160603,Gopalakrishnan:2018wd,PhysRevLett.122.127202,10.21468/SciPostPhys.6.4.049,Bulchandani_2021,Gopalakrishnan_2023}.  \par
	One can understand the origin of this diffusion as coming from two ingredients: ballistic motion of the quasiparticles (magnons and boundstates thereof) and fluctuations in the quasiparticle’s charge which is ``screened'' by the magnetization fluctuations of the initial state. That is to say if in a region of size $\ell$, the magnetization density fluctuations scale as $\ell^{-w}$, where $w$ is known as the wandering exponent. Then, if the quasiparticle travels a distance $\ell$ the fluctuations of the quasiparticle’s charge over that distance also scales as $\ell^{-w}$.\par
	For initial states drawn from a thermal ensemble, these fluctuations obey the central limit theorem ($w=1/2$) and using this fact along with the ballistic motion of the quasiparticles one can show that this leads to diffusive spin transport~\cite{PhysRevLett.122.127202}. This suggests that tuning the wandering exponent of the initial state should modify the nature of spin transport.\par
	In this work we demonstrate that this is indeed the case by constructing an ensemble of initial states with tunable wandering exponent which have fluctuations that grow faster than a thermal ensemble's. Based on the charge screening argument outlined above, we show that these states should display superdiffusive transport and indeed find that this is the case by numerically computing the variance of the charge transfer. The variance of the charge transfer grows in a power law fashion and provide an argument for the expected exponent of this power law. We find relatively good agreement with numerical results and explain how deviations arise from finite size effects.\par

\begin{figure}[!t]
	\includegraphics[width=\columnwidth]{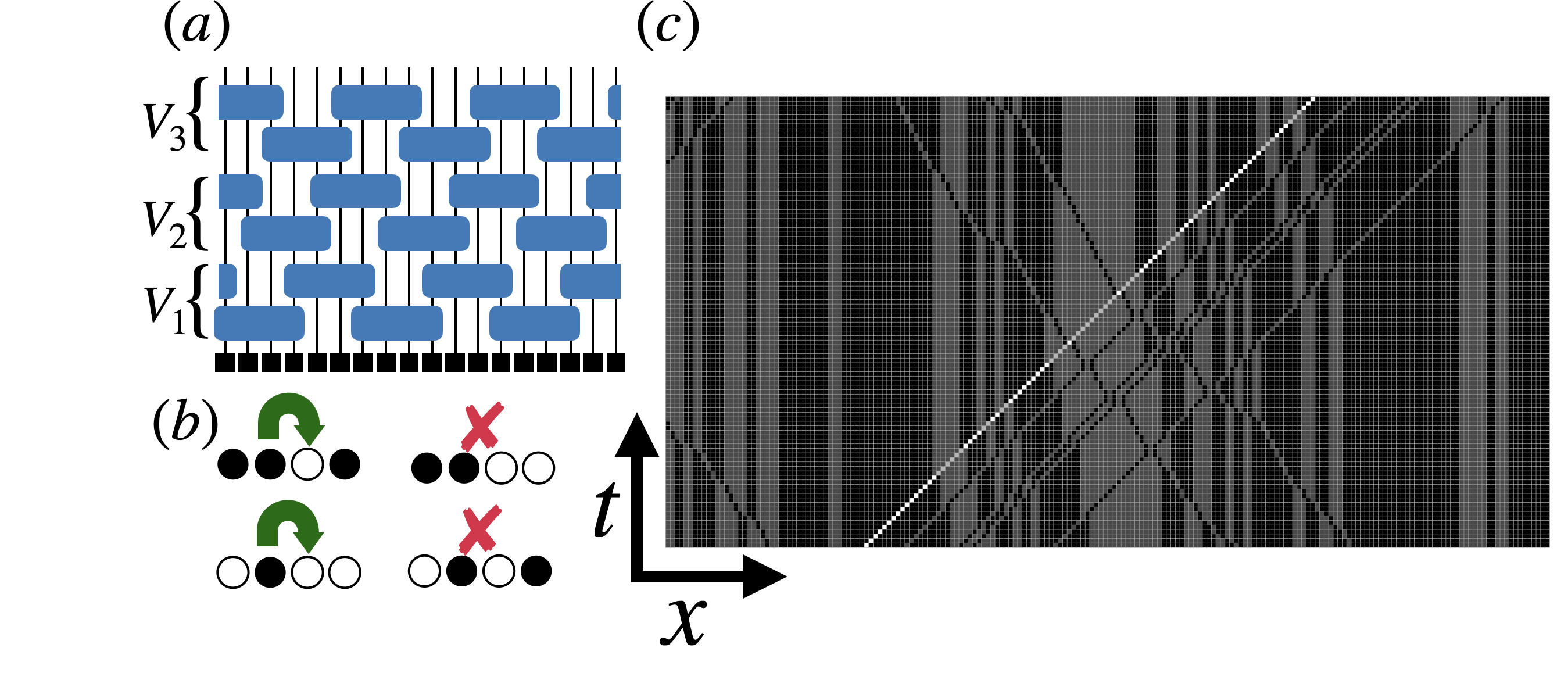}
	\caption{{\bf Folded XXZ Automaton.} (a) Circuit geometry for the XXZ automaton follows a staircase pattern. (b) The rules for the updates in the automaton. One can see that they conserve the number of domain walls. (c) A highlighted magnon trajectory traversing ballistically through domains.}
	\label{fig:fig1}
\end{figure}
\emph{Screening Argument}.--- To illustrate our argument, we study transport in the folded XXZ automaton \cite{Pozsgay_2021} which qualitatively captures transport features of the easy axis regime and can be thought of as the $\Delta \rightarrow \infty$ limit of the quantum spin-1/2 XXZ chain~\cite{10.21468/SciPostPhysCore.4.2.010,10.21468/SciPostPhys.10.5.099}. A desirable feature of this model compared to the spin-1/2 XXZ chain is that product states in the occupation basis are mapped to product states and thus we can perform simulations to long times with large system sizes. Our argument will rely solely on a quasiparticle picture, and we expect it to generalize to the entire diffusive regime $|\Delta|>1$ of the XXZ spin chain. \par
 The system is comprised of $L$ qubits whose individual basis states are $|\bullet\rangle$ (particle) and $|\circ\rangle$ (hole). The unitary governing the dynamics is given by $\mathcal{U}=V_3 V_2 V_1$ where $V_{j} = \prod_{i\equiv j\!\!\!\mod 3}U_{i,i+1,i+2,i+3}$ and
\begin{equation} 
\begin{split}
U_{i,i+1,i+2,i+3} &= P^{\bullet}_i  \mathrm{SWAP}_{i+1,i+2} P^{\bullet}_{i+3}\\
&+P^{\circ}_i  \mathrm{SWAP}_{i+1,i+2} P^{\circ}_{i+3}\\
&+P^{\circ}_{i}P^{\bullet}_{i+3}+P^{\bullet}_{i}P^{\circ}_{i+3},
\end{split}
\end{equation}
where $P^{\circ} = |\circ\rangle\langle\circ| $ and $P^{\bullet} = |\bullet\rangle\langle\bullet| $, conserving both particle and domain wall number.  A pictorial representation of the circuit along with the update rule performed by the gates are shown in Fig.~\ref{fig:fig1}. This model is integrable and has three types of quasiparticles: left- and right-moving magnons (single isolated particles or holes), and frozen domains which arise from domain wall conservation. \par
To understand transport in integrable systems we essentially have to keep track of how much charge has been spread by the magnons. To track the amount of charge a magnon spreads we compute $\delta x_{\mathrm{charge}}(t)^2 = \langle (q_{\mathrm{mag}}(t) x_{\mathrm{mag}}(t))^2 \rangle $ where $q_{\mathrm{mag}}$ denotes the charge of the magnon, $x_{\mathrm{mag}}$ denotes the distance the magnon has traveled, and $\langle \cdot \rangle$ denotes averaging over the ensemble of initial statess. In integrable systems magnons traverse ballistically with a well-defined velocity, $v$ (we imagine tracking a single species), so
\begin{equation}
\delta x_{\mathrm{charge}}(t)^2 = v^2t^2 \langle q_{\mathrm{mag}}(t)^2 \rangle.
\end{equation}
On average the magnon's charge will be zero at infinite temperature but the variance will non-zero. As shown in Fig.~\ref{fig:fig1}, magnons change their charge by passing through frozen domains. For example, if a magnon is a particle prior to collision, it will change to a hole while traversing the next domain (see Fig.~\ref{fig:fig1}). This means a magnon's charge fluctuations are entirely specified by the pattern of frozen domains in the initial state. Fluctuations in the pattern of frozen domains are determined by the initial state magnetization density, $m$, defined such that a particle/hole has charge $\pm 1$. Thus, fluctuations are characterized by the wandering exponent, $w$, i.e. in a region of size $\ell$, $m \sim \ell^{-w}$.\par 
Thus we have that the charge of the magnon should scale in the same way as the initial state magnetization density fluctuations. So if a magnon moves a distance $\ell$ in a time $t$ then $\langle q^2_{\mathrm{mag}} \rangle \sim \ell^{-w} \sim t^{-w}$. Therefore
\begin{equation}
\delta x_{\mathrm{charge}}(t) \sim t^{1-w}.
\end{equation}\par
This argument was first used in Ref.~\cite{PhysRevLett.122.127202} and was applied to thermal states where $w=1/2$ where one correctly predicts diffusive spin transport. In the next section we will provide a procedure to generate states with tunable wandering exponent with $w<1$ which implies superdiffusive spin transport.
 \par 
\begin{figure*}[!t]
	\centering
	\includegraphics[width=\textwidth]{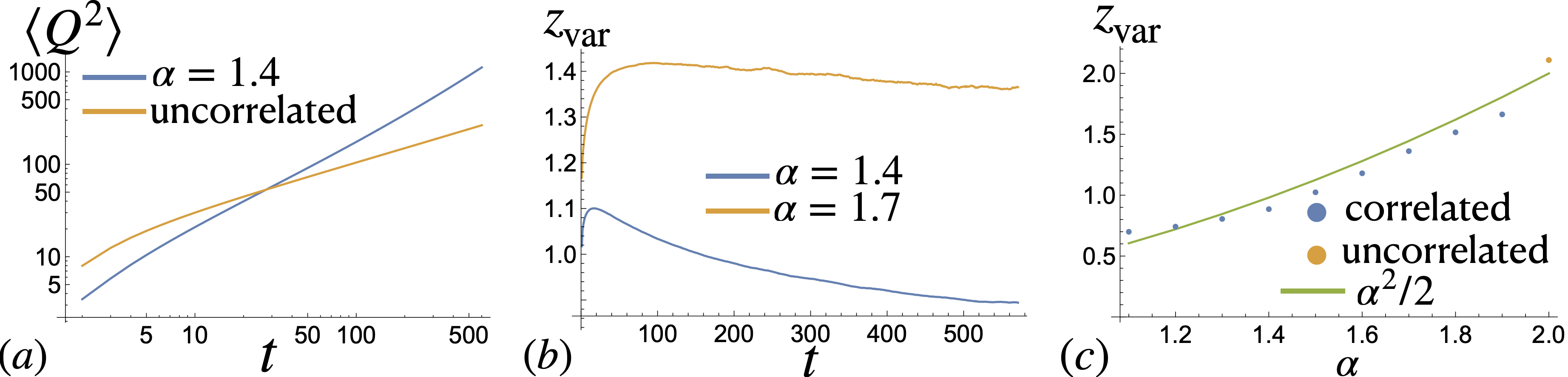}
	\caption{{\bf Charge Variance vs Stability Parameter.} (a) Variance of the charge transfer as a function of time shown for stability parameter, $\alpha =1.4$, compared to initial states drawn from an infinite temperature, i.e. uncorrelated, ensemble. Note that the fluctuations for $\alpha=1.4$ grow faster than those in the thermal ensemble indicating superdiffusive behavior. (b) Dynamical exponents, $z_{\mathrm{var}}$, versus time obtained via log derivatives for two different values of $\alpha$. (c) $z_{\rm var}$ obtained via log-derivatives at a late time, $t=500$. The orange point $\alpha=2$ is obtained by sampling states from an infinite temperature ensemble. One sees that the prediction, $z_{\mathrm{var}} = \alpha^2/2$ is roughly consistent with the numerics. The data is averaged over $10^7$ initial states.}
	\label{fig:fig2}
\end{figure*}
\emph{Correlated States}.---
Based on the screening argument, generating states which do not obey the central limit theorem should lead to anomalous transport. One way to do so is to create states that are spatially correlated. This can be achieved by forming product states which are comprised of contiguous fully polarized domains (e.g. $|\bullet\bullet\bullet\bullet\circ\circ\circ\circ\bullet\bullet\bullet\circ\circ\rangle$) that have unbounded lengths. To do so we draw the lengths of these domains from L\'evy alpha stable distributions (although any distribution with power law tails should work)~\cite{BOUCHAUD1990127}. These are distributions with power law tails that can be tuned via the so-called stability parameter that characterizes the distribution. To be more precise, let $S(y; \alpha,c,\beta,\mu)$ denote the stable distribution with stability parameter, $\alpha$;  scale parameter, $c$; skewness, $\beta$; mean, $\mu$.\par 
 In this work $\mu=0$ and $\beta=0$ and $1<\alpha\leq 2$. For large $y$, $S(y)\sim y^{-(\alpha+1)}$. Given a random variable $y$ distributed according to $S$ then the domain length of the $i$th domain is given by $x_i = \mathrm{floor}(|y|)$. For $1<\alpha\leq2$ the average of $y$ is well defined and we note that the average of $|y|$ for $\beta=\mu=0$ is  given by $\frac{2}{\pi}c^{1/\alpha} \Gamma(1-1/\alpha)$ where $\Gamma(z)$ is the usual gamma function and so the scale parameter determines the average domain size. We randomly choose a domain to be fully filled (empty) with equal probability.  \par 
To determine the wandering exponent of these states we compute the magnetization ((un)occupied sites have charge $\pm 1$)  in a region of size $\ell$, i.e. $M= \sum_{i=1}^\ell \sigma_i$ and $\sigma_i=\pm 1$. Let $\tau_i$ denote the polarization of the $i$th domain (this is $\pm1$ for occupied and unoccupied respectively). We can rewrite the sum as $M = \sum_{i=1}^k x_i \tau_i + (\ell - \sum_{i=1}^{k} x_i)\tau_{k+1}$ where the last term accounts for the fact not all $k$ domains may fit inside the region. Let the magnetization of a domain be $m_i = x_i \tau_i$. Then the distribution of $m_i$ is symmetric since $\tau_i=\pm 1$ with equal probability and has power law tails inherited from the domain length distribution. Thus we can invoke the generalized central limit theorem to show that $\langle M^2 \rangle \sim k^{2/\alpha}$. For the stability parameters we consider, the mean is well-defined and so the typical domain lengths will be controlled by $c$. For $c\sim O(1)$, we expect that the number of domains $k\sim \ell$. Thus $\langle M^2 \rangle \sim \ell^{2/\alpha}$. We conclude that the wandering exponent is given by $w=1-1/\alpha$.\par 
For $1< \alpha \leq 2$, this means that magnetization fluctuations decay slower than thermal states consistent with the fact that these states have larger correlations. Based on the screening argument this would imply that charge spreads as $t^{1/\alpha}$ which would indicate superdiffusive transport. In the next section we demonstrate that this is indeed the case by computing transport observables. \par
%
\emph{Transport and numerics}.---
To characterize transport in this special class of initial states, we cannot use the standard linear-reponse Kubo formalism (which relies on special properties of thermal states). Instead, we diagnose charge transport via the variance of the so-called charge transfer $Q(t)$,
\begin{equation}
Q(t) = \sum_{x\leq 0 } (n_{x}(t)-n_{x}(0))- \sum_{x> 0} (n_{x}(t)-n_{x}(0)),
\end{equation}
with $n_x = 0,1$ the particle density on site $x$.
The charge transfer keeps track of how much charge has been transferred across the origin at time $t$. The full distribution of the charge transfer is known as the full counting statistics and it not only captures linear response, but also captures details about fluctuations on top of the average hydrodynamic behavior~\cite{Levitov1993,Levitov1996ElectronCS,Ivanov1997,Belzig2001,Belzig2002,Levitov_2004,Lesovik13,Krajnik:2022aa,Krajnik:2022ab,Krajnik:2022ac, Krajnik:2023aa,Gopalakrishnan:2022aa,McCulloch:2023aa}. Experimental setups can access the full distribution via snapshots corresponding to projective measurements on the whole system~\cite{Mazurenko2017,Gross2017,Bakr_2009,FCSSchmiedmayer,Weitenberg_2011,Bohrdt_2021,Parsons_2016,Hilker_2017,Mitra_2017,Haller2015,Sherson2010,bloch2012,Bernien_2017,Islam_2011,Zhang_2017,Garttner_2017,Song_2017,GoogleSupremacy,Blais_2021,Wendin_2017,doi:10.1126/science.abk2397}. In this work, we will only be focusing on the behavior of the variance, $\langle Q^2 \rangle - \langle Q \rangle^2$, for states at half filling, i.e. $\mu = 0$. When $\mu = 0$, $\langle Q \rangle = 0$ but $\langle Q^2 \rangle \neq 0$ and is generally expected to grow in a power law fashion, $\langle Q^2 \rangle \sim t^{1/z_{\mathrm{var}}}$ and $z_{\mathrm{var}}=2$ for diffusive systems. In this section we numerically compute $\langle Q^2 \rangle$ and extract $z_{\mathrm{var}}$ via a log-derivative, i.e. $z_{\mathrm{var}} (t) = 2 \bigg(\frac{d\log \langle Q^2 \rangle }{d \log t} \bigg)^{-1}$.\par 
Our results are shown in Fig.~\ref{fig:fig2}. One can see that when initial states are drawn from the correlated ensemble $\langle Q^2 \rangle$ grows faster than initial states drawn from a thermal ensemble at infinite temperature, which is consistent with the screening argument. We extracted $z_{\mathrm{var}}$ for various $\alpha$ and one can clearly see that tuning $\alpha$ does indeed change the behavior of $\langle Q^2\rangle$ which is consistent with the fact that tuning fluctuations of the initial state should correspond to tuning transport. \par 

\emph{Theory of charge transfer}.---
To understand the behavior of $Q(t)$ we first consider initial states where the magnetization density $m_x = (2 n_x-1)/2$ for $x\leq 0$ ($x > 0$) is $m/2$ ($-m/2$) and the imbalance, $m$, is large. In this limit one has a low density of magnons in each half of the chain so there is a well-defined domain wall that separates the two halves of the chain. For example an initial state might look like 
\begin{center}
\parfig{.2}{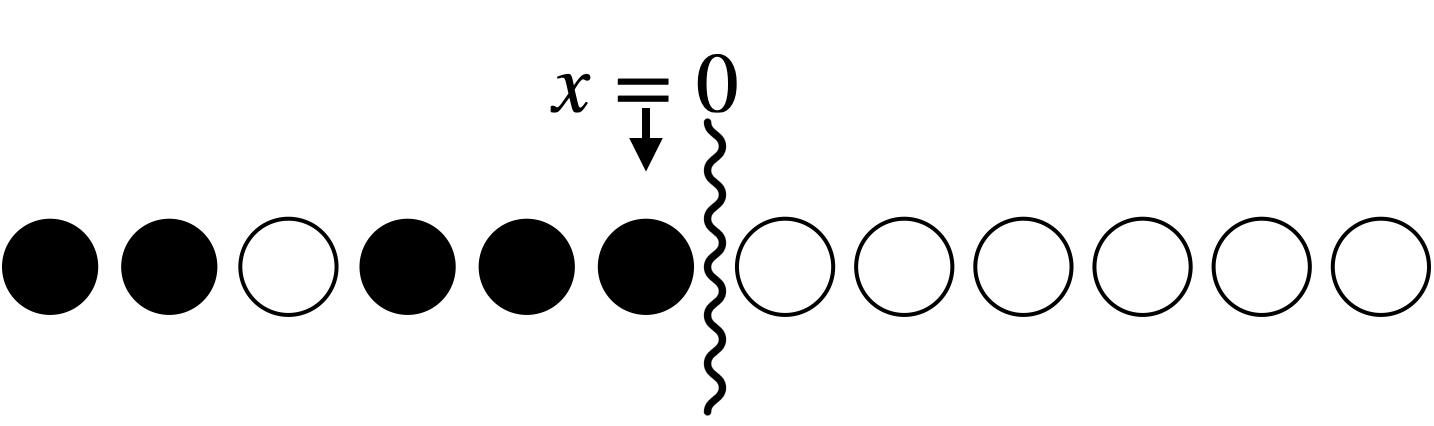},
\end{center}
 where the squiggly line indicates the location of the domain wall which separates the halves of the chain.  Due to domain wall conservation the only time charge can be transferred is if a particle or hole moves across the origin. At some later time the magnon will cross the origin and our state will look like 
 \begin{center}
\parfig{.2}{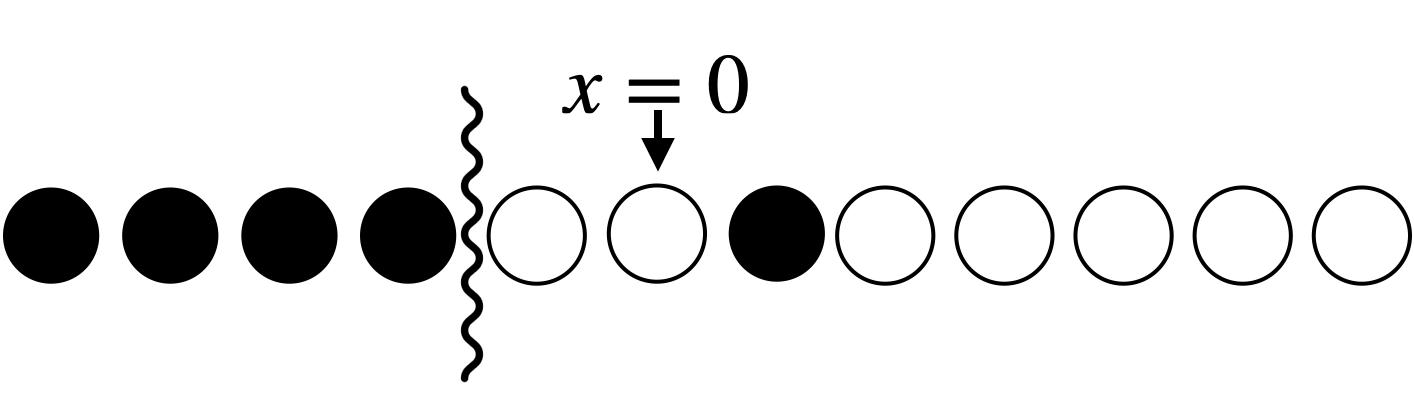}. 
\end{center}
 Notice that the domain wall that separated the halves of the chain moves two sites in the opposite direction in which the magnon moved. The key observation that was made in Ref.~\cite{Gopalakrishnan:2022aa} is that charge transfer is linked to the motion of this domain wall that separates the two halves. If at time $t$, the domain wall is at $x_{\mathrm{dw}}(t)$ and if $m_{[0,x_{\mathrm{dw}}]}$ is the magnetization density in the region between the origin and the location of the domain wall then
\begin{equation}
Q(t) = -2m_{[0, x_{\mathrm{dw}}(t)]}x_{\mathrm{dw}}(t).
\label{eqn:cteqn}
\end{equation}\par
When $m\rightarrow 0$ there is not a well defined notion of a domain wall that separates the two halves of the chain. However one can imagine a fictitious domain wall which is located at the origin and moves in the same fashion as one would expect when $m\simeq 1$. For instance, suppose we have the following initial state, 
 \begin{center}
\parfig{.2}{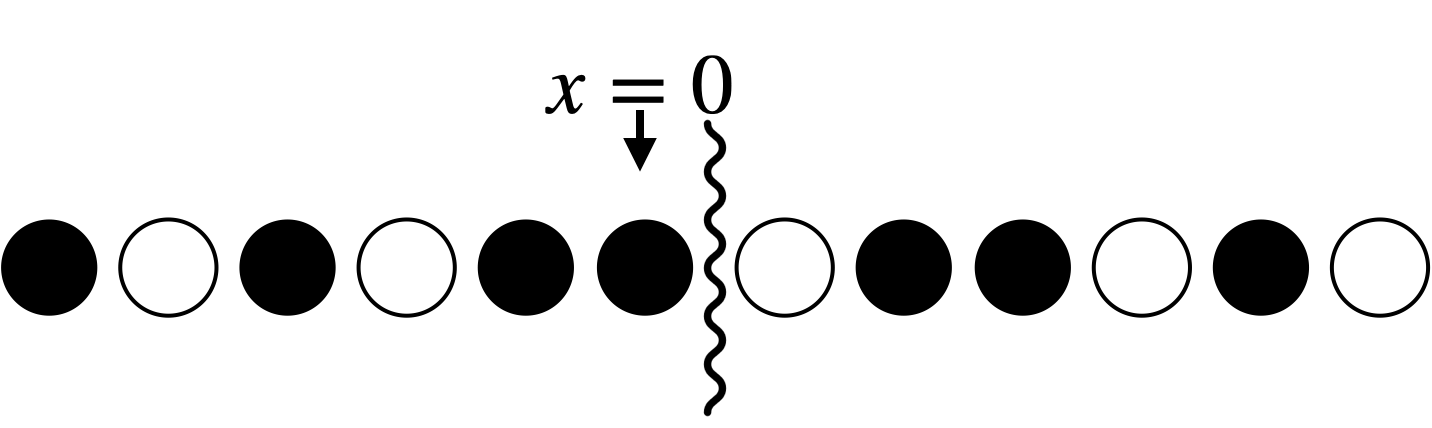}, 
\end{center}
 where the squiggly line indicates the location of our fictitious domain wall initially at the origin.  At the next time step the hole next to the origin crosses over to the left half of the system and our state will look like 
 \begin{center}
\parfig{.2}{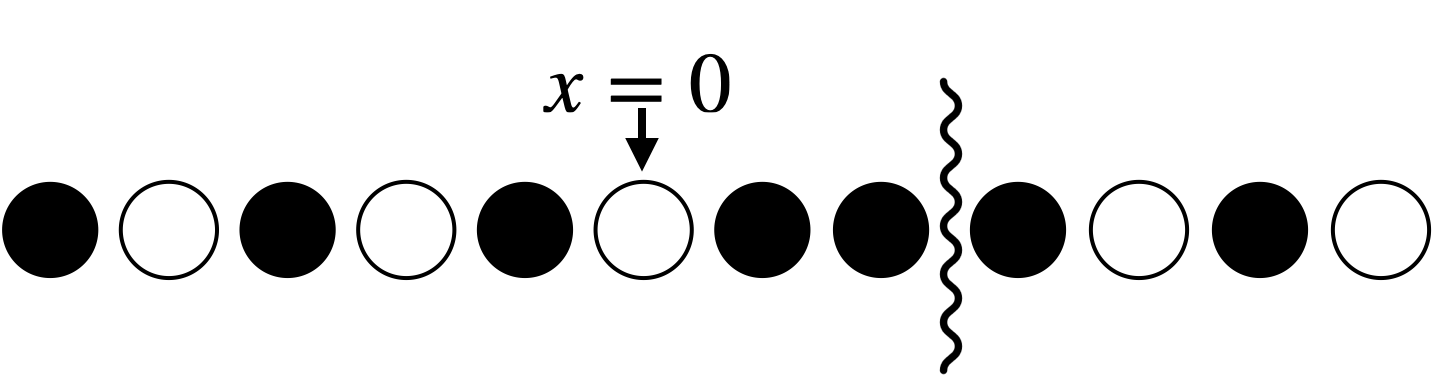}. 
\end{center}
 The domain wall moves two steps in the opposite direction in which the magnon propagated. Demanding that the fictitious domain wall move in this manner implies that the charge transfer satisfies Eq.~\ref{eqn:cteqn}. \par 
From Eq.~\ref{eqn:cteqn}, we can now relate the dynamical exponent of the charge transfer variance to the motion of the domain wall. On average the domain wall will be at the origin at half-filling, i.e. $\langle x_{\mathrm{dw}}(t) \rangle = 0$, since the number of left and right movers is equal, but $\langle x_{\mathrm{dw}}(t)^2 \rangle \neq 0$ and is expected to grow as a power law, $\langle x_{\mathrm{dw}}(t)^2\rangle\sim t^{2/z_{\mathrm{dw}}}$. Using the fact that the magnetization density over a distance $|x_{\mathrm{dw}}|$ scales $m\sim |x_{\mathrm{dw}}|^{-1+1/\alpha}$, then the variance of the charge transfer should obey the following scaling relation,
\begin{equation}
\langle Q(t)^2\rangle\sim t^{2/\alpha z_{\mathrm{dw}}}.
\label{eqn:ctscaling}
\end{equation}\par 
 \begin{figure}[!t]
	\includegraphics[width=\columnwidth]{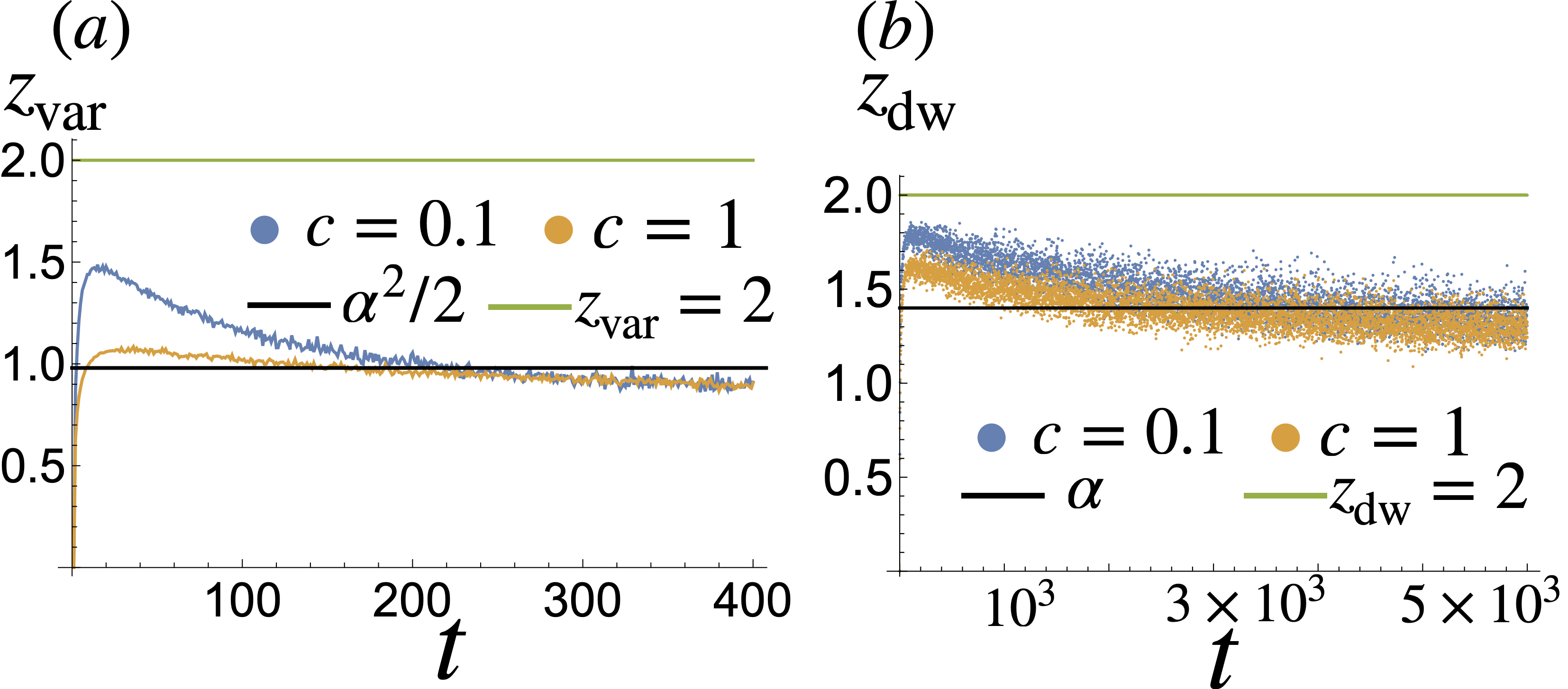}
	\caption{{\bf Dynamical Exponent Finite Size Effects.} (a) One sees that at short times the dynamical exponent of the variance of the charge differs for different scaling parameters but at longer times converges to the same value. Here data was averaged $10^7$ initial states. (b) The domain wall dynamical exponent as function of time features much stronger finite sizes as the exponent does not appear to converge until $\sim 10^3$ which is an order of magnitude larger than the variance of the charge transfer. Here results were averaged over $10^6$ realizations.}
	\label{fig:fig3}
\end{figure}
To obtain what $z_{\mathrm{dw}}$ is we note that the domain wall only moves when a magnon ``hits" it. And whenever the magnon ``hits" the domain wall its charge changes thus the variance in time of $x_{\mathrm{dw}}$ should follow the same time dependence as the variance in the amount of charge a magnon has carried, i.e. $\delta x_{\mathrm{mag}}(t)$. Since $\delta x_{\mathrm{mag}}^2\sim t^{2-2w} = t^{2/\alpha} $ then $\langle x_{\mathrm{dw}}(t)^2\rangle  \sim t^{2/\alpha}$. Using this result along with Eq.~\ref{eqn:ctscaling} implies $z_{\mathrm{var}} = \alpha^2/2$. From Fig.~\ref{fig:fig2}, we see relatively good agreement with our prediction. \par  
Numerical deviations from this prediction can be understood from two finite size/time effects: the motion of the domain wall and the dependence of the dynamical exponents on $c$. The first comes from the fact that in a time $t$ the domain wall will have only moved a distance $t^{1/\alpha}$ and thus the domain wall cannot efficiently probe the tails of the distribution. The second comes from the fact that the mean domain size of is tuned by the scaling parameter, $c$, and at long times and distances one expects the dependence on $c$ to drop out and only the tails of the distributions should matter. As shown in Fig.~\ref{fig:fig3}, we see that $z_{\mathrm{var}}$ and $z_{\mathrm{dw}}$ display a dependence on the scale parameter with the more severe dependence in the latter case presumably coming from the fact that the domain wall does not probe the tails of the distribution well enough and thus is much more prone to experience stronger effects from the scaling parameter. \par
\emph{Discussion}.--- In this work we demonstrated that anisotropic integrable spin chains provide a route towards tunable transport by tuning initial state magnetization fluctuations. We gave a procedure to generate states which have quasi-long range correlations using L\'evy alpha stable distributions. And we demonstrated that the fluctuations in this ensemble of correlated states is tuned by the stability parameter that changes the tails of these stable distributions. \par
Using the variance of the charge transfer we found superdiffusive transport in these initial states which is tuned by the stability parameter consistent with the screening argument set forth in Ref.~\cite{PhysRevLett.122.127202}. Furthermore we are able to provide a theoretical prediction for the dynamical exponent controlling the growth of the charge variance via the motion of a “domain wall” and the magnetization density in the region between the position of the domain wall and the position at which the charge transfer is being measured. We find good agreement with our prediction and numerical simulations with some finite size effects.\par
We showed that these finite size effects come about from the domain wall’s inability to efficiently probe the tails of these stable distributions at short times. As such the domain wall motion is severely affected by the average domain size that it probes which is controlled by the scaling parameter of the L\'evy alpha stable distributions. Surprisingly the time scale for which the dependence on $c$ goes away is different for the charge variance and the domain wall motion. This suggests that finite size effects in the magnetization density cancel out the finite size effects in the motion of the “domain wall.” It would be interesting to find different ensembles of correlated states that are able to probe the tails more effectively. \par
While we only presented results for superdiffusive transport in principle one can also achieve tunable subdiffusive transport. For this to happen one would need to generate states which have fluctuations that drop off faster than those satisfying the central limit theorem. Such states, also known as hyperuniform states, would need to have domains that are anti-correlated.\par
A second interesting route would be to go away from the $\Delta \rightarrow \infty$ limit and probe transport in the $\Delta > 1$ regime of the spin-1/2 XXZ chain using these correlated initial states. In this regime fully polarized domains are no longer frozen but are only frozen for a time $t\sim \Delta^{s-1}$ where $s$ is the domain length. The dynamics of the spin-1/2 XXZ chain is accessible in experiments~\cite{XXZhelices} and so it would also be interesting to explore this tunable transport experimentally since these correlated initial states are product states. Approaching the isotropic $\Delta=1$ limit should be particularly interesting, as it already shows superdiffusive transport with $z=3/2$ for thermal states~\cite{Znidaric:2011aa,Ljubotina:2019aa,Bulchandani_2021, 2023arXiv230515463G,doi:10.1126/science.abk2397}.   \par
 \emph{Acknowledgements}.---
We thank Vedika Khemani for stimulating discussions. This work was supported by the National Science Foundation under NSF Grants No. DMR-2103938  (S.G.), DMR-2104141 (R.V.), DMR-1945529 (M.K.), the Welch Foundation through award number AT-2036-20200401 (M.K.), and the Alfred P. Sloan Foundation through a Sloan Research Fellowship (R.V.) .

\bibliography{corr_refs,refs}

\end{document}